\documentclass{article}
\usepackage{graphicx}
\usepackage{amssymb,amsmath}

\usepackage[UKenglish]{babel}
\usepackage[margin=3.0cm]{geometry}
\usepackage{float}
\usepackage[hypertexnames=false]{hyperref}
\usepackage[noend]{algpseudocode}
\usepackage{algorithm}
\usepackage{amsmath,amssymb,mathtools}
\usepackage{amsthm}
\usepackage{cleveref}
\usepackage[toc,page]{appendix}
\usepackage{siunitx}
\usepackage{cite}
\usepackage{pgfplots}
\usepackage{subcaption}
\usepackage{chngpage}
\pgfplotsset{compat=1.7}

\usepackage{authblk}

\DeclareGraphicsExtensions{.eps,.jpg,.pdf,.ps}

\begin{document}
\date{}
\title{Propagation in rough waveguides: Forward and inverse problems} 

\author[1]{M. Spivack\thanks{\url{ms100@cam.ac.uk}}}
\author[1]{O. Rath Spivack\thanks{\url{or100@cam.ac.uk}}}

\affil[1]{Department of Applied Mathematics and Theoretical Physics,
  University of Cambridge, UK}

\font\eightrm=cmr8

\maketitle

%\ead{or100@cam.ac.uk}

{\eightrm
\begin{center}
$^1${Department of Applied Mathematics \& Theoretical Physics,\\
                  Centre for Mathematical Sciences,
                      University of Cambridge,
                  Cambridge CB3 0WA, UK
}
\end{center}
}

\begin{abstract}

We discuss here the direct and inverse problems for wave propagation in
a waveguide with rough internal surface and arbitrary mean shape. The
high degree of multiple 
scattering inside the waveguide poses significant challenges both for the
forward computation and for the recovery of the surface
profile, and raises important questions about scattering cross-sections.
This paper falls into two parts corresponding to these issues.

We first apply Left-Right (L-R) operator splitting to calculate the
scattered fields in 2- and 3-dimensional waveguides. Using this we
illustrate the scattered  
fields and the effect of surface roughness.
In the second part we formulate an algorithm for surface recovery from field
measurements along the waveguide axis, which generalises recent work on
surfaces in 2 dimensions.  This method utilizes forward scattering
assumptions in effect by 
formulating an integral equation in the unknown surface field, treated
as a function of the surface.  Although discussed in the context of
waveguides,  the formulae are given in a
form applicable to a variety of geometries, with coefficients
which will depend on and be determined by each specific application.

\end{abstract}

\def\del{\partial}
\def\beq{\begin{equation}}
\def\eeq{\end{equation}}
\def\beqn{\begin{eqnarray}}
\def\eeqn{\end{eqnarray}}
\def\rbar{{\bf r}}
\def\x{{\bf x}}
\def\b{{\bf b}}
\def\H{{\bf H}}
\def\Jinc{{\J_{inc}}}
\def\JJ{{\J_{0}}}
\def\JK{{\tilde\J_{inc}}}
\def\C{{\bf C}}
\def\D{{\bf D}}
\def\J{{\bf J}}
\def\t{{\bf t}}
\def\n{{\bf n}}
\def\A{{\cal A}}
\def\L{{\cal L}}
\def\Linv{\L^{-1}}
\def\R{{\cal R}}
\def\u{{\bf u}}
\def\v{{\bf v}}
\def\w{{\bf w}}
\def\f{{\bf f}}
\def\g{{\bf g}}
\def\P{{\bf P}}
\def\s{{\bf s}}
\def\p{{\bf p}}
\def\S{{\bf S}}
\def\U{{\bf U}}
\def\V{{\bf V}}
\def\W{{\bf W}}
\def\intf{\int_{-\infty}^{\infty}}
\def\intfs{\int_{{\rm surface}}}

\section{Introduction}

Imaging of rough surfaces from acoustic or electromagnetic scattering
data is a significant challenge in a wide range of applications,
attracting a great deal of attention in various regimes using a
variety of approaches
\cite{wombell,bao2016,yapar,cayoren,akduman,zhang2017,warnick, 
  zhang1,zhang2,sefer2024imaging,sefer2020iterative,SR2}. 
However, fewer results are available for rough ducts or
waveguides, despite their importance in fields such as biomedical
imaging, non-destructive testing, and detection of low cross-section aircraft.
The scarcity of results is largely due to the computational and analytical
difficulties inherent in calculating fields within extended
arbitrarily curved waveguides.

We consider here propagation in a waveguide with a rough internal surface
and given arbitrary mean shape, in both 2D and 3D. The study falls into
two distinct parts: the forward problem for which we implement an efficient
operator splitting approximation previously applied to smooth-sided
waveguides; and the inverse problem of profile recovery.
%for which a formulation is given but not implemented here.
For the latter problem, we give, without implementing here, a general
formulation which is applicable to a range of geometries having a
principal direction of propagation.
This reformulates the marching algorithm which has been validated
in previous work on 1-dimensional rough surface recovery
\cite{SR2,chen2018rough,chen2018rough2}.  The algorithm is expressed,
for each unknown surface coordinate, 
as a closed form expression depending explicitly on upstream values.
The formula includes coefficients which are to be determined in the
specific regime in which it is applied.  Applications will be
presented in subsequent work. 

Both of these tasks present significant
difficulties due to severe computational complexity and a high degreeo
f multiple scattering.  However, the predominance of forwardscattering can be turned to advantage; Left-Right operator splitting
(closely related to the Method of Multiple Ordered Interactions and
the Forward-Backward method) proves accurate and efficient and is
well-suited to this purpose
\cite{tran1997calculation,kapp1,pino,spivack2017efficient}.

For the forward problem we calculate scattering for a range of
parameters along canonical curved waveguides, with and without surface
roughness, in both two and three dimensions. We use this to examine
the scattered signal and the effect on this of roughness.
Preliminary results are given for individual realisations and
ensemble averages.  The key features are captured by
taking just two terms of the series solution; in fact backscatter
is entirely due to the second.  The results are intended to be
illustrative, and a much wider study is required especially
three-dimensions to properly explore cross-polarisation and 
variation with parameters.

For surface reconstruction using measurements along the duct axis, the
proposed algorithm casts the problem as an integral equation in the
unknown surface field, treated as a function of the surface, and
extends recent work on the recovery of an near-planar 1D surface
\cite{chen2018rough,chen2018rough2}.  As mentioned above this yields a
an expression for each unknown surface coordinate depending explicitly
on upstream values, with coefficients to be determined depending on
the application.

This paper is organised as follows: The forward problem for a
rough-sided 2-dimensional duct is treated in section \ref{forward2d}
in terms of coupled integral equations following the L-R operator
series method of
\cite{spivack2001validation,spivack2002electromagnetic}. The numerical
solution is summarised and computational examples given for an
elongated rough duct.  In section \ref{forward3d} the generalisation
to a 3-dimensional rough duct is shown, giving the series solution and
example calculations.  Finally the inverse problem is discussed in
section \ref{inverse_sec}, and the marching or `surface sweep'
solution is presented.

\section{Propagation in rough waveguides: forward calculation}

\subsection{Rough two-dimensional waveguide\label{forward2d}} 

  In this section we consider the propagating fields due to a wave
  impinging on a (possibly curved) rough-sided 2-dimensional
  waveguide.
  The formulation is expressed throughout terms of
  electromagnetic waves but in this setting reduces to a scalar
  problem and is equally applicable to acoustics.

  respectively. 
\subsubsection{Formulation}
 The two bounding surfaces are perfectly conducting and denoted $S_1$ and $S_2$
 For convenience we summarise here the governing
  equations and approach from \cite{spivack2002electromagnetic} for a
  smooth canonical waveguide.

The aperture, $A$, is at the left. 
The incident field on this aperture may be  a horizontally
or vertically polarized plane wave (TE or TM, equivalent to Dirichlet
or Neumann boundary  conditions for an acoustic wave).  
The field internal to the waveguide is written as 
a boundary integral over the field induced by the aperture field along
its surface. We will assume vertical polarization so the  
magnetic field $H$ obeys the Helmholtz wave equation $(\nabla^2+k^2)H=0$.  
Let $G$ be the free space  Green's function, so that $G$ is the zero
order Hankel function of the first kind,
\beq 
G(\rbar,\rbar') = {1\over 4i} H_0^{(1)} (k|\rbar-\rbar'|) .
\eeq
where $\rbar=(x,z)$ and $x$ and $z$ denote
horizontal and vertical coordinates respectively.

The forward problem is to obtain the surface currents along the
duct, and from these to calculate the internal fields.
The field at a point $\rbar$ inside the waveguide is written
as a boundary integral along the duct surfaces:
\beq
H_s({\bf r}) = H_{inc}({\bf r}) + \int_S
~\left[ {\partial G(\rbar,\rbar') \over\partial \n} H
({\rbar'})  \right] ~ dr'
\label{w5}
\eeq
where $\n$ denotes the normal outward from the waveguide,
 $\rbar_s=(x,S(x))$, $\rbar'=(x',S(x'))$
and $S$ is the union of the two surfaces.
Taking limits as $\rbar$ tends to the upper and to the lower
surfaces and applying boundary conditions, we obtain coupled integral
equations in $H$ along the two surfaces.
It is convenient to regard these as separate functions of the single coordinate
$x$:
\beq
%\begin{eqnarray}
  {J}_1 (x) = H (x,S_1(x))~, ~~~~~~~~~~~~~~ {J}_2 (x) = H  (x,S_2(x)) .   
\label{w6} 
%\nonumber
%\end{eqnarray}
\eeq
The coupled integral equations for $ {J}_1$, $ {J}_2$ can then be written
\begin{eqnarray}
 H_{inc}({\bf r_1}) &=&
\int_S \left[ G_\n({\bf r_1};{\bf r_2'}) {J}_2(x') -
G_\n({\bf r_1};{\bf r_1'}) {J}_1(x') \right] dr' 
\nonumber \\
H_{inc}({\bf r_2}) &=&
\int_S \left[ G_\n({\bf r_2};{\bf r_2'}) {J}_2(x') -
G_\n({\bf r_2};{\bf r_1'}) {J}_1(x') \right] dr' 
\label{w7} \\ \nonumber
\end{eqnarray}
where $G_\n$ denotes the normal derivative of $G$, and 
${\bf r_i}=(x,S_i(x)),~~~{\bf r_i'}=(x',S_i(x'))$ for $i=1,2$.
This set of coupled equations can be solved to find
the field terms $ {J}_1$, $ {J}_2$ along the surfaces, which
may be substituted into
equation (\ref{w5}) to yield the value of the field in the waveguide.

\medbreak
For a given external field impinging on the duct, the incident field $H_{inc}$
can be obtained as an integral over the
aperture.  This represents the field which
would exist beyond the aperture, in the absence of the waveguide itself.
Here the external field is taken to be a plane wave.

\bigbreak
For convenience we now write eq.(\ref{w7}) in operator notation,
\begin{eqnarray}
 H_{inc}({\bf r_1}) &=&
(L_{11}+R_{11}) {J}_1 + (L_{12}+R_{12}) {J}_2
\nonumber \\ 
 H_{inc}({\bf r_2})  &=& 
(L_{21}+R_{21}) {J}_1 + (L_{22}+R_{22}) {J}_2
\label{3.2} \\ \nonumber
\end{eqnarray}
where the operators are defined by

\begin{eqnarray}
L_{11} f &= 
{1\over 2}  f - 
\int_0^x  G_\n ({\bf r_1; r_1'}) f(\rbar'_1)
dr_1', ~~~~~~~~~~~%\nonumber \\
R_{11} f &= - 
\int_x^X  G_\n ({\bf r_1; r_1'}) f(\rbar'_1)
dr_1', \nonumber \\
L_{12} f &=
- \int_0^x G_\n ({\bf r_1; r_2'}) f(\rbar'_2)
dr_2', ~~~~~~~~~~~%\nonumber \\ 
R_{12} f &=
- \int_x^X G_\n ({\bf r_1; r_2'}) f(\rbar'_2)
dr_2' \label{4.5} \\ \nonumber
\end{eqnarray}
with the obvious definitions for $L_{21}$, $L_{22}$, $R_{21}$, $R_{22}$.
In these expressions the left-hand integrals $L_{ii}$ are 
interpreted as the Cauchy principal value, and $r_i'$ indicates integration 
along surface $S_i$.

\def\u{{\bf u}}
\def\f{{\bf f}}
Eqs. (\ref{3.2}) can be expressed more concisely if we
define the vectors $\u$ and $\f$ as the pairs of functions 
\beq %\begin{eqnarray}
\u= \left({J}_1, {J}_2 \right), ~~~~~~~~~~
\f = \left(H_{inc}(\rbar_1), H_{inc}(\rbar_2)\right) .
\eeq %\end{eqnarray}
We can then write (\ref{3.2}) in the form 
\def\J{{\bf J}}
\def\K{{\bf K}}
\def\L{{\bf L}}
\def\R{{\bf R}}
\def\A{{\bf A}}
\beq 
\f ~~=~~ \A \u ~~\equiv~~ (\L + \R) \u 
\label{4.6}
\eeq
where  $\L$ and $\R$ are the $2\times 2$ matrix operators whose entries are 
integral operators $L_{ij}$, $R_{ij}$ corresponding to the above
splitting (see ref []).
$\L$ therefore accounts for scattering from 
the left. The formal solution is given by 

\beq  
\u = (\L+\R)^{-1}\f . \label{4.7} 
\eeq
The main task is the inversion of $(\L+\R)$.  
As in previous treatments [12] we make use of the assumption
that the effect of $\R$ is `small' in some sense,
reflecting the fact that most of the scattering is due to interaction 
from the left.  The solution (\ref{4.7}) can be then expanded in a series
\beq
\u = \left[1 - \L^{-1} \R + ( \L^{-1} \R)^2 - ...\right] \L^{-1} \f 
\label{4.8}
\eeq
Provided it converges this equation can be truncated and treated term by term.
When the system is discretized, the operator $\L$ becomes
a lower triangular matrix operator, whose entries are simply the four
lower-triangular matrices arising from $L_{ij}$.
Similarly $\R$ becomes a upper triangular matrix operator in which the
matrices are strictly upper triangular. 
Inversion of the matrix $\L$ can be carried out very 
efficiently (using Gaussian elimination and backward substitution)
to give the first term of eq.(\ref{4.8}).
Since subsequent terms in the series are products of $\L^{-1}$ and $\R$, they
can also be evaluated efficiently.

\vskip 2 true cm

\subsubsection{Numerical Solution}

The main steps in the numerical treatment of the equations are to
truncate the series, discretize the integral equation, and solve the
resulting linear system.  
The details of the discretization itself are identical to those set
out elsewhere  \cite{spivack2002electromagnetic}  and need only
outlined briefly. 

\medbreak
The series solution
(\ref{4.8}) can be regarded as an iterative solution, of which one or two terms 
are sufficient (see e.g. \cite{spivack2001validation}).
The first term gives say
$\u_1 =  \L^{-1} \f $. % \label{(3.14)} 
The first correction $\u_2 = \L^{-1} (\R u_1) $ to this is calculated
by applying the  integral operator $\R$ and inverting $\L$ again, to give 

$$ \u \cong \u_1 - \u_2 . $$
This gives the required surface currents along the waveguide
boundaries, which is then used as the driving field for the calculation of
interior fields, using eq. (\ref{w5}). To discretize we divide the
integrals into evenly spaced subintervals, using around 10 points per
wavelength.  Each subintegral can be  
linearise and expressed approximately as a coefficient times the
unknown function value. As an example, the $L_{11}$ term in equation
(\ref{4.5}) can be written 
as a matrix equation, say
\beq 
H_{inc} = A H . \label{(3.11)} 
\eeq
in which the matrix $A$ has entries
\begin{eqnarray}
A_{mn} &=& {i\delta\over 4}\sigma_n
\left.{\del H_0^{(1)}\over \del n}\right|_{kr_{mn}}  ~~~~~{\rm for}~~
m\neq n \nonumber\\
A_{m m} &=& - \left[
{1\over 2} - {\delta \over\sigma_m^2 \pi}s''(x_m) \right] \label{(3.12)} \\
\nonumber
\end{eqnarray}
where $\sigma_n = \sqrt{1+s'(x_n)^2}$ is the slope variable
where the prime denotes the $x$-derivative, $h'=dh/dx$, and
the off-diagonal entries $m\neq n$ have simply approximated
by the Green's function, with a factor accounting 
for arc-length. 

\medbreak 
\subsubsection{Computational Results}

The waveguide profile is specified by the addition of a continuous
stochastic process $r(x)$, say, to a smooth (canonical) surface.  The
canonical shape chosen here is one for which the truncated series
solution has been carefully validated
\cite{spivack2001validation}. Surface variation $r(x)$ is Gaussian
distributed and is generated with chosen variance $\sigma^2$ (or
r.m.s. height $\sigma$) and given autocorrelation function $\rho(\xi)
= <r(x)r(x+\xi)>$.  The length scale of the variation along the duct
wall can be characterised by the distance over which $\rho$ decays by
a factor of $e^{-1}$. (The roughness is statistically stationary so
that $\rho$ is a function of spatial separation $\xi$ only.)

\medbreak

\begin{figure} %[h]
\includegraphics[width=0.5\linewidth]{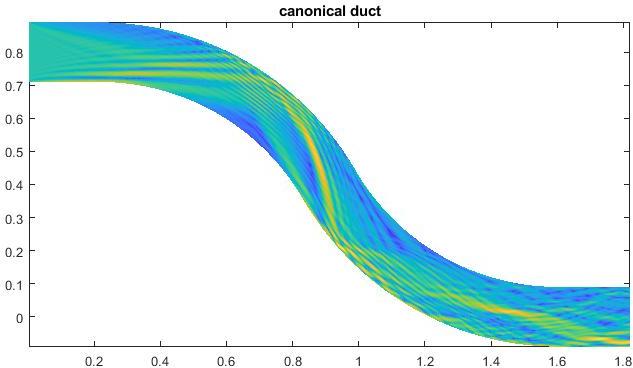}
	\hskip 0.2 true cm    
	\includegraphics[width=0.5\linewidth]{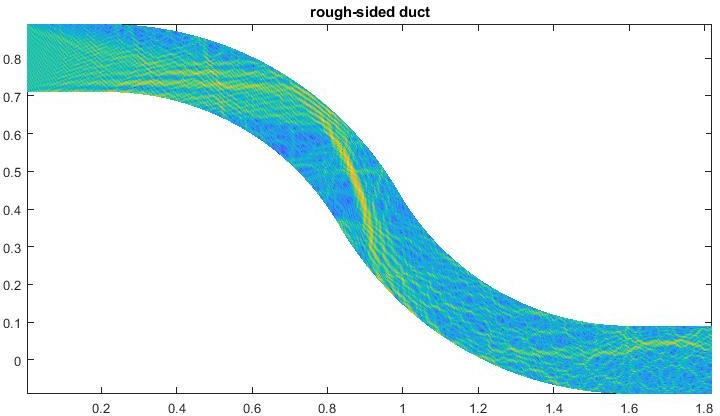}
	\caption{Canonical duct (left) compared with slightly rough
          duct, fine scale roughness (right)}  
	\label{fig1} 
\end{figure}

\medbreak In the examples shown in Figure \ref{fig1} the aperture $A$
is at the left, the depth at $A$ is $20\lambda$, and the overall
length with respect to $x$ is $200\lambda$, where $\lambda$ is the
wavelength. A plane wave is incident on the aperture, at an angle of
$10^o$ to the horizontal. This induces a surface current, and the
resulting field amplitude internal to the duct is shown.  The
left-hand figure shows the smooth-sided duct and on the right is the
rough-sided duct. In this case the scale size is around 0.01.
It is seen here that even a small perturbation $r$ leads to
significant degradation of the pattern which would otherwise be
observed.  This highlights the importance of realistic modelling for
such problems.  It is this perturbation which allows reconstruction of
the surface roughness profile.

\begin{figure} [ht]
	\hskip 4.0 true cm    
	\includegraphics[width=0.4\linewidth]{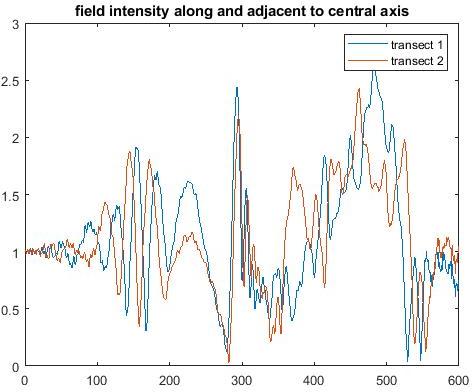}
	\caption{Measured field intensity along two curves adjacent to
          the canonical duct  
	axis for the example of Fig. \ref{fig1}} 
	\label{fig3} 
\end{figure}

Figure \ref{fig3} shows the field intensity along two curves running
along the duct adjacent and just above and below the axis.

\medbreak
We now consider the backscattered field or scattering cross-section,
the variation with  
incident angle, and the effect of surface roughness upon this. Note
that this is  
illustrative and not intended to be a full exploration of parameter space.

The backscatter fields are calculated as follows: The scattered
component $H_s$ of the  
field is obtained in the same way as the previous figures. The
backscattered field in the  
region $x<0$ results entirely from $H_s$ across the aperture, since
the remaining field  
$Hinc$ is composed entirely of right-travelling waves.     We can
therefore calculate the  
external scattering cross-section by evaluating the
Helmholtz-Kirchhoff integral of $H_s$  
at the aperture.  (This is analogous to the procedure applied to
obtain the incident  
field itself beyond the aperture.)   

\begin{figure} %[h]
	\centering 	
		\includegraphics[width=0.7\linewidth]{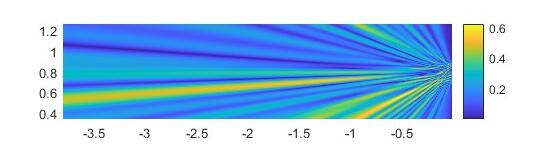}
		\caption{Backscatter due to $0^o$ incident fields,
                  rough duct: single realisation} 
		\label{fig10b} 
\end{figure}

\begin{figure} %[h]
       \centering
		\includegraphics[width=0.6\linewidth]{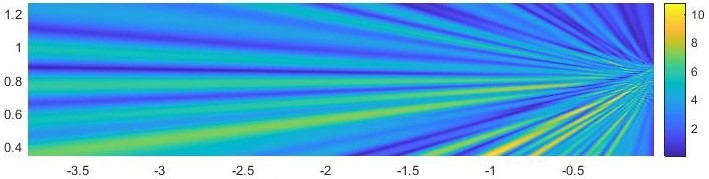}
		\caption{Backscatter due to $0^o$ incident fields,
                  rough duct: sum over 40 realisations} 
		\label{fig10d} 
\end{figure}

\begin{figure} %[h]
	\centering
	\includegraphics[width=0.7\linewidth]{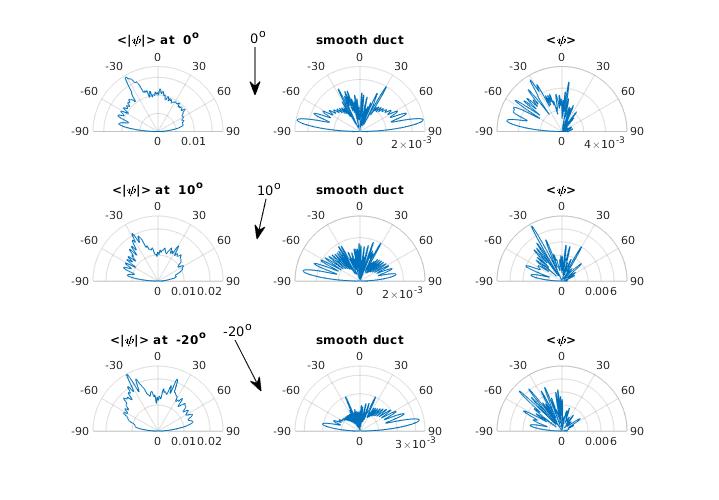}
	\caption{Directivity patterns for 3 incident angles, averaged over 50 
	realisations of rough duct showing mean amplitude (left), mean
	field (right), and corresponding pattern for zero roughness (middle).}  
	\label{directivity} 
\end{figure}

Figures \ref{fig10b} to \ref{fig10d} first show the spatial intensity
pattern, including the  
near field, in order to illustrate the roughness effects, for a field
incident normal to  
the aperture,  The amplitude pattern has been scaled with distance
from the aperture   
plane by applying a $1/\sqrt{x}$ factor to account for geometrical spreading. 
Figures 
\ref{fig10b}, \ref{fig10d} show respectively a single realisation and
the sum over multiple  
realisations.  A striking feature which is seen here is the
asymmetrical `ray' in both  
coherent field and mean amplitude. 
In \ref{directivity} we show the directivity pattern for a range of
incident angles  
averaged over multiple realisations, compared side-by-side with the
smooth duct case. 
These were obtained by first calculating the field and normal
derivative at the aperture,  
and applying a far-field Helmholtz-Kirchhoff integration.  It is
interesting to note the  
persistent peak which occurs for all incident angles, but only in the
presence of roughness, at around $120^o$, that is $40^o$ from normal to the
aperture.   This peak
%was not observed occur for cylindrical ducts, and
may be regarded as a signature of the waveguide geometry which is enhanced by
the roughness.

Recall that these calculations were done using the first two terms of
the L-R series to produce the surface currents, followed by an `exact' numerical
surface integral. If only the first L-R term is used, the surface
integral (roughly speaking) includes  backscattered rays which do not
interact with the surface; in other words it gives direct  
backscatter only.    On the other hand, the second term allows for
from constructive interference between reversible ray paths; such
paths cannot occur for direct backscatter (apart from specular).

We now consider a linear waveguide of varying cross-section.

\begin{figure} [ht]
	\centering
	\includegraphics[width=0.7\linewidth]{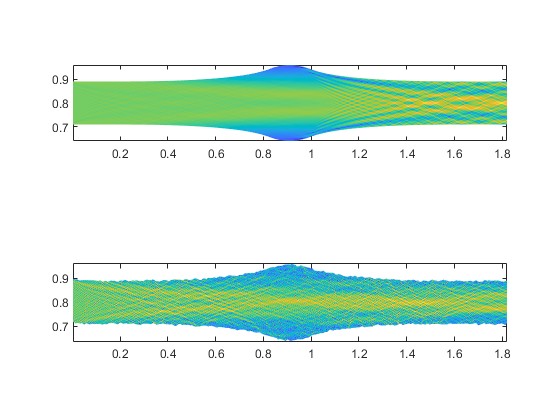}
	\caption{Intensity pattern in a smooth-sided and rough-sided waveguides of varying
          cross-section (respectively upper and lower plots, at zero
          degrees incidence}
	\label{bulbous0deg} 
\end{figure}

\begin{figure} %[h]
	\centering
	\includegraphics[width=0.7\linewidth]{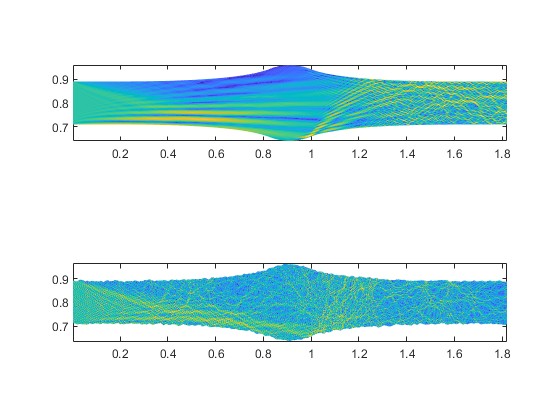}
	\caption{Intensity pattern in a smooth-sided and rough-sided waveguides of varying
          cross-section (respectively upper and lower plots, at a $10^o$
          angle of incidence directed downwards} 
	\label{bulbous10deg} 
\end{figure}

\bigbreak

\subsection{Forward problem: Rough three-dimensional waveguide
\label{forward3d}}

The formal operator splittng solution described above
generalises relatively straightforwardly to 3-dimensional ducts.  In
principle, in fact, it is more straightforward than in 2-dimensions
since the coupled integral equations are replaced by a single,
although higher-dimensional, integral equation.  The main complication
is in the numerical treatment, since there is a further horizontal
direction transverse to the propagation direction; the lower
triangular matrices of the previous section now become
block-lower-triangular. Consequently, at {\sl each} step in the
marching procedure a matrix inversion is required. However, this is
relatively fast, since that matrix dimension corresponds to the number
of points around the duct circumference, rather then the much larger
total number of unknowns.  This has been described elsewhere. We
describe this briefly in this section and give some numerical results.
 
For the perfectly conducting (PEC) case, the governing Stratton-Chu 
or magnetic field integral equations [MFIE] must
be solved over the irregular boundary to obtain electric surface current $\J$;
this surface current is then used in a scattering
integral to obtain the electromagnetic signature.  

\beq
\J_{inc} = {1\over 2} \J - 
\n\! \times\! \! \! \intfs \! \! \! \! {\hskip -.3 true cm}
\J\! \times\! \nabla G ~dS  
\eeq
We again tackle this problem using the Left-Right
Splitting method (L-R), in which successive terms model increasing orders of
multiple scattering.  

We again write the above equation formally as
$\A{\J} = {\Jinc}$, 
where ${\Jinc}$ is
the field incident (say) from the left, so that 
 ${\J}=\A^{-1}{\Jinc}$.  
The inverse of $A$ can again formally be expressed as a series
\beq 
\A^{-1}= \L^{-1} - \L^{-1} \R \L^{-1} + ...
\label{eq1} \eeq 
Discretization of the integral equation yields a block
matrix equation,  in which $\L$ is the lower triangular part of 
the block matrix 
$\A$ (including the diagonal) and $\R$ is the upper triangular part.
Under the assumption that most energy is right-going, $\L$ is the
dominant part of $\       A$, and the series can be truncated to
provide an approximation for ${\J}$.   This approach has several
advantages. In terms of wavelength $\lambda$, 
evaluation of each term scales with the fourth rather than the 
sixth power of $\lambda$ required for $\A^{-1}$.
Subsequent terms, of which typically only the first one or two are needed,
each have the same computational cost.  With further approximations this 
can be reduced to $\lambda^3$. However, the low complexity and memory
requirement allow very large problems to be tackled without such
additional approximations.  
The algorithm also lends itself well to parallelisation, and
the speed scales approximately linearly with the number of processors.

\begin{figure} %[h]
  \centering
  \begin{minipage}{.5\textwidth}
  \centering
%	\hskip 2.0 true cm    
	\includegraphics[width=0.8\linewidth]{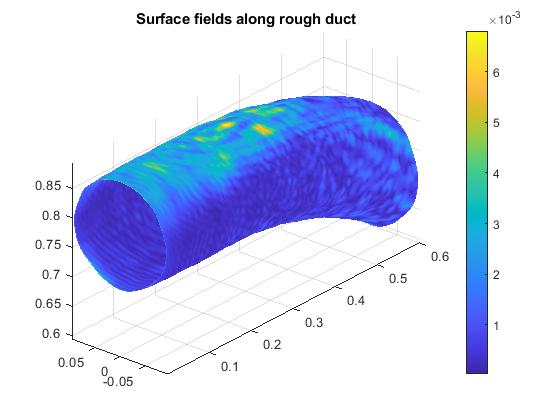}
	\caption{Slightly rough duct of length $34\lambda$} 
	\label{fig2b} 
%\end{figure}
%\begin{figure} [h]
  \end{minipage}%
\begin{minipage}{.5\textwidth}
  \centering
%	\hskip 2.0 true cm    	
\includegraphics[width=0.8\linewidth]{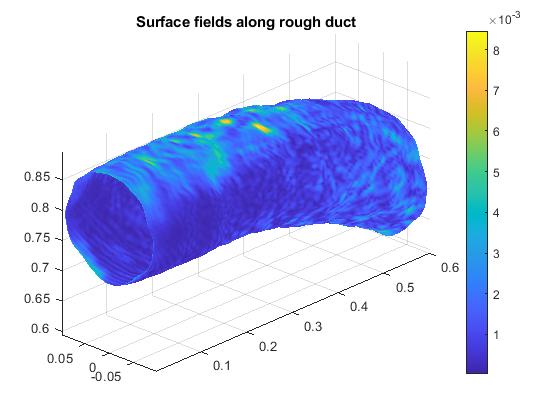}
	\caption{Rough duct, length $34\lambda$} 
	\label{fig3b} 
    \end{minipage}
\end{figure}

\subsubsection{Numerical Results}
We present some examples for the 3-dimensional rough-sided waveguide.
Figure \ref{fig2b} is a short section of around 34 wavelengths of a
curved duct with moderate roughness. The aperture is 20 wavelengths in
diameter and is illuminated by a vertically polarised 10GHz plane wave
at $10^o$ from the horizontal. The scattered component of the surface
current amplitude induced by the incident field is shown. In  
Figure \ref{fig3b} is an example of the duct with more significant
roughness, for the same incident field and aperture size. The section
shown in both cases is around 34 wavelengths, and colour represents
the surface current amplitude. 

\begin{figure} %[h]
  \centering
  \begin{minipage}{.5\textwidth}
  \centering
%	\hskip 2.0 true cm    
%\vskip 0.4 true cm
	\includegraphics[width=0.7\linewidth]{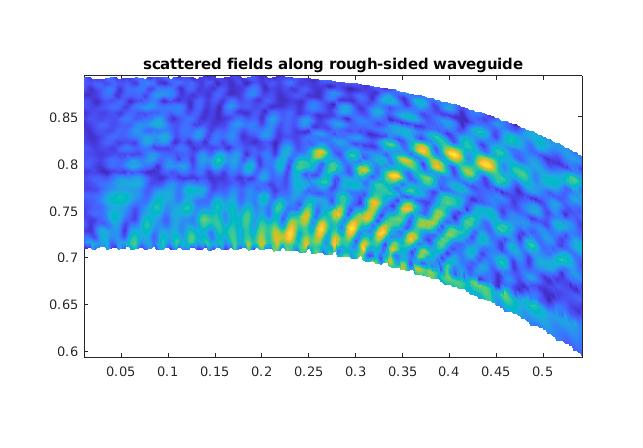}
	\caption{Duct length $34\lambda$, 10GHz} 
	\label{fig6} 
     \end{minipage}%
  \begin{minipage}{.5\textwidth}
  \centering
%\end{figure}
%\begin{figure} [h]
%	\hskip 2.0 true cm    
	\includegraphics[width=0.7\linewidth]{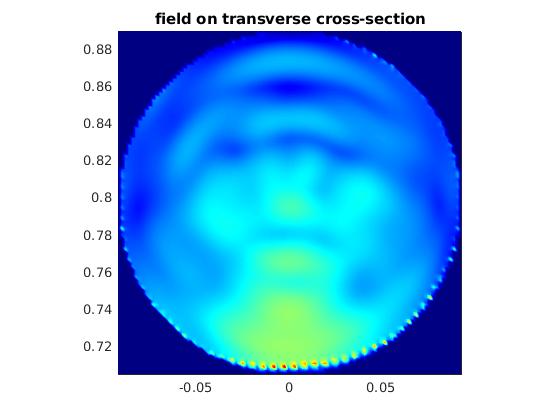}
	\caption{Duct length $34\lambda$, 10GHz} 
	\label{fig7} 
     \end{minipage}%
\end{figure}

.

\begin{figure} 
	\hskip 4.0 true cm    
	\includegraphics[width=0.6\linewidth]{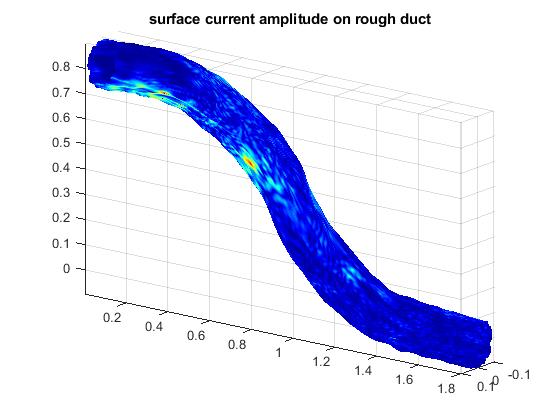}
	\caption{Surface current amplitude at 32GHz along rough duct} 
	\label{fig8} 
\end{figure}

\medbreak For this latter case, the resulting fields internal to the
duct were calculated. Figure \ref{fig6} shows the field intensity
along a vertical plane running along the middle of the duct, i.e. the
intersection of the $(x,z)$-plane with the duct.  The field on a
cross-section transverse to the propagation direction, i.e. parallel
to the $(y,z)$-axis, is given in Figure \ref{fig7},

Finally, shown in Figure \ref{fig8}, we calculate the surface currents
on a larger duct of around 200 wavelengths, with the fields resulting
from a 32GHz plane wave incident on the aperture.

\bigbreak

\bigbreak

\section{Inverse problem: Reconstruction of waveguide
  surface\label{inverse_sec}}

\medbreak 

In this section we set out the principles of the
marching method inverse solution which has been applied elsewhere, and
formulated here inm the context of an irregular waveguide.  
The intentiona is to express the algorithm in compact explicit form,
which may be applied in multiple regimes, and which we expect to be
amenable to convergence and regularisation analysis.  The algorithm is
not implemented here; this will be carried out in future work.

The approach is motivated by earlier work and is
largely based on forward scattering assumptions and L-R operator
expansion. Although we describe key quantities in terms of a
waveguide, equations are derived in a more general setting in order to
encompass various 2D and 3D geometries.  The specific details
for each implementation will depend closely on the geometry and the
Green's function in each case.
 
\medbreak 
Suppose $\s = \s(x,\phi)$ is the duct surface, where $x$ is
the horizontal component of the propagation direction and $\phi$ is a
radial (or more generally transverse) coordinate.  We will suppose
that the underlying smooth (canonical) duct shape is known.   We aim
to recover the surface profile $\s(x,\phi)$ from the equations relating unknown
surface fields (both incident and scattered) to the measured data.
The main properties on which the method is based are as follows: The
forward-scattering nature of the approximation allows us to
reconstruct the solution sequentially, sweeping along the propagation
direction $x$.  In 2D settings \cite{spivack2001validation} this has
been found to be both 
efficient and remarkably robust, exhibiting a type of
self-regularisation with repsect to measurement noise.  The incident field at 
$\s(x,\phi)$, depends only on $\s(x,\phi)$ itself; this allows use of
an invertible transformation within the algorithm.  The unknown surface field
is approximated by its dependence on `upstream' values, and crucially
its dependence on 'local' surface values is weak (in a sense to be
discussed).

\medbreak Let $\w(x)$ represent the scattered field measured along the
central axis of the duct.  At each $x$ this will consist of a vector
of $N_\phi$ values where $N_\phi$ is the number of points we wish to
recapture around the duct at each distance $x$, which for simplicity
will be assumed constant in $x$.  We will assume a known underlying
`canonical' duct shape. 
(For an extended surface varying about a plane $N_\phi$ would
represent the number of points transverse to the propagation
direction. For a rough 2D duct, $N_\phi=2$, and for a rough surface in
2D $N_\phi=1$.)

\medbreak Let $\u(x,\phi)$ be the unknown surface field. When
discretized this will again become a vector, of size $N_\phi$ for each
$x=x_i$, where $i=1,..,n$.  The $x$-values are again assumed evenly
spaced; this is for convenience and is not an inherent limitation of
the proposed algorithm.  Let $\v(\s(x,\phi))$ be the incident field
along the surface, considered as a function of $x$.
Applying the series solution to the coupled integrals and taking 
only the first term, i.e. using $\A\cong\L$, $\A^{-1}\cong\Linv$ gives
\beq
\begin{split}
\u (x,\phi) ~&\cong~ \Linv \v \\
\w(x,\phi)  ~&\cong~ \L\u
\end{split}
\label{inteq}
\eeq

All terms here including the operators depend on the unknown rough 
profile.   We will suppose that the duct aperture $\s(0,\phi)$ is known.

\medbreak
Denote by $\S, \U, \W, \V$ the discretized forms
corresponding respectively to $\s$, 
$\u$,  $\w$, and  $\v$, over the whole computational domain, so that
$\U, \W$ and $\V$ all depend on $\S$. 
Denote by $\S_i, \U_i, \W_i, \V_i$  the vectors  of length $N_\phi$
evaluated at distance $x_i$.  
%\medbreak   
In equations (\ref{inteq}), $\u(x,\phi), \w(x,\phi)$ and operators $\Linv, 
\L$ evaluated at $x$ depend only on upstream  values $\s(X \le x,\phi')$  
so we can write
%\medbreak

\beqn
%\begin{split}
%\begin{aligned}
	\U_i ~&=~ \sum_{j=1}^i M_{ij} \V_j  \nonumber\\
	\W_i  ~&=~ \sum_{j=1}^i  N{ij} \U_j 
%\end{aligned}
%\end{split}
\label{inteq2}
\eeqn
where $M_{ij}$ and $N_{ij}$ are transformations derived from the operators in 
(\ref{inteq}).  
Suppose that at step $x_i$ we have already obtained the surface values
at all range  
steps 
$x_j$ for $j < i$.  We therefore rearrange this to obtain coupled 
equations
\beqn
%\begin{split}
%\begin{aligned}
\U_i - M_{ii}\V_i~&=~ \sum_{j=1}^{i-1} M_{ij} \V_j \nonumber  \\
\W_i - N_{ii} \U_i ~&=~ \sum_{j=1}^{i-1}  N_{ij} \U_j 
%\end{aligned}
%\end{split}
\label{inteq3}
\eeqn
in which the left-hand-side is unknown, while the 
terms on the right are known.   The quantities $M_{ii}, N_{ii}$ are
are $N_\phi$-dimensional matrices.  We have thus reduced the problem to that of
expressing the LHS explicitly in terms of the, as yet unknown, values
$\S_i\cong  
s(x, \centerdot)$.

\medbreak
The above considerations and equations (\ref{inteq3})
encompass the inverse problem for the 3D duct and the rough surface in 3D,  
the 2D duct ($N_\phi=2$) and the rough surface in 2D ($N_\phi=1$).  
It is instructive to revisit the 2D problem, for which a
similar approach was used: Consider an irregular 
1-dimensional surface.  
Here the scattered field $w$ is measured along a line parallel to the
mean plane. To  
simplify notation we rewrite this as:
\beq
\begin{aligned}
u_i - \alpha_{i}v_i~&=~ \sum_{j=1}^{i-1} \alpha_{ij} v_j 
\\
w_i - \beta_{i} u_i ~&=~ \sum_{j=1}^{i-1}  \beta_{ij} u_j 
\end{aligned}
\label{inteq4b}
\eeq
Each term on the left %(\ref{inteq4b})
is now a scalar.  The values $\alpha_{i}$ and $\beta_{i}$, depend in a
known way on the surface but it is found
\cite{Direct,spivack1990moments}
that they may be approximated by their values at the canonical flat
surface.  The measured data point is $w_i$, and the incident field $v_i$
can be approximated linearly in terms of the unknown surface height
$s(x_i)$, say $v_i=\gamma_i s_i$.  This results in a pair of equations
in the two unknowns $u_i, s_i$ which we can write as \beq
\begin{aligned}
u_i - \alpha_{i}\gamma_i s_i ~&=~ \sum_{j=1}^{i-1} \alpha_{ij} v_j 
%\label{inteq5a}
\\
\beta_{i} u_i             ~&=~ w_i ~-~\sum_{j=1}^{i-1}  \beta_{ij} u_j  
\end{aligned}
\label{inteq5b}
\eeq

This system is overdetermined since $u_i$ depends on $s_i$ and in
previous work this issue has been circumvented is various ways
\cite{SR2,chen2018rough2}  .  Here we initially treat $u$ and $s$ as
locally independent, and note that second equation (\ref{inteq5b})
contains only $u_i$; we can therefore solve for $u_i$ and substitute
into the upper equation (\ref{inteq5b}) to find $s_i$.  [To summarise
  the rationale for this: $u_i$ depends on $s_j$ for \textit{all}
  $j\le i$; the dependence of $u_i$ `locally' on $s_i$ is therefore
  weak, and $s_i$ can be replaced by its canonical or mean surface
  value at $x_i$.]

\medbreak
We thus obtain the marching solution:
\beq
s_i = \frac{-1}{\alpha_{i}\gamma_i}
\left\{  \sum_{j=1}^{i-1} \alpha_{ij} v_j - \frac{1}{\beta_i}\left[
 w_i ~-~\sum_{j=1}^{i-1}  \beta_{ij} u_j
\right]
\right\}
\label{inverse}
\eeq This gives the solution in the case of a rough surface varying
about a mean plane in 2D.  It can be refined by iterative improvement
to take into account the interdependence between surface profile and
the resulting wavefields. An initial guess or assumption
is needed for the value $s_1$, but the system has been found very
robust with respect to this choice.  The forward scatter
assumption can also be relaxed, although this is less straightforward
as it is highly regime-dependent.

\medbreak Returning to the more general case, an analogous
higher-dimensional procedure can be applied: In equation
(\ref{inverse}), the term $\gamma_i$ remains a scalar. However,
$\alpha_i$ and $\beta_i$  become $N_\phi\times N_\phi$
matrices; the reciprocals become matrix inverses, to be evaluated at
each step $x_i$.  This yields:

\beq
\S_i = \frac{-1}{\gamma_i}  M_{ii}^{-1}
\left\{  \sum_{j=1}^{i-1} M_{ij} \V_j -  N_{ii}^{-1}\left[
\W_i ~-~\sum_{j=1}^{i-1}   N_{ij} \U_j
\right]
\right\}
\label{inverse3D}
\eeq

\medbreak It should be emphasized that the dimensions of matrices to
be inverted here are relatlvely small, so that the algorithm is
computationally tractable. Indeed the computational cost is broadly
comparable with that of the L-R series for the forward problem.

\medbreak The above formulation assumed that data is measured and
surface reconstruction is required at the same (evenly-spaced) values
of $x$.  It was also assumed implicitly that the data were arranged
parallel to the surface; that is along the duct axis, or on a plane
parallel to the mean plane in the extended surface case.  Neither of
these assumptions is an inherent limitation and both can be relaxed,
provided the number of available data points is not reduced.  However,
since the scattered wave is forward-going, the optimal measurement
location with respect to distance from the surface depends on the
angle and spread of the incident wave.  This issue was not critical in
previous applications because of the robustness and self-regularising
characteristics with respect to measurement noise, but it should
ideally be taken into account.

%\hrule

\section{Conclusions}

In this paper we have, first, calculated wave scattering inside
rough-sided ducts, and both 2- and 3-dimensional example geometries, by the
operator expansion method of Left-Right splitting.  This utilizes
the predominance of forward scatter but allows the calculation of
multiple-scattering backscatter.  For the 2-dimensional case we have
examined mean directivity patterns for various angles. These patterns
arise from the second term of the L-R series which captures
the dominant multiple backscattering contributions.

\medbreak
We then derived a marching solution of the inverse problem,
that is the recovery of the surface profile from measured data. This
was expressed in terms of the waveguide, but is applicable for a range of
geometries having a principal wave propagation direction,
This marching or `surface sweep' method
recovers the surface values sequentially at successive points along
the surface.  The approach makes use of the first approximation to the
L-R series solution, and is an explicit method, apart from
coefficients which will depend on the specific geometry.  This is not
implemented here. The application to particular cases including the
waveguide will be given in future work.

\medbreak
It may be observed that this raises the possibility of tackling an important 
dual inverse problem: the design of a boundary profile for a given
incident field which will produce a given reflectivity pattern. 

\medbreak
A long-term goal is to solve the surface reconstruction problem for a
three-dimensional lined duct. In such geometries, even the  
forward problem, which is required for almost all inverse
calculations, can be problematic because of prohibitive computational
costs.  This problem wll be treated in future work based on the fast
forward solver used here.

\vskip 2 true cm
\section*{Acknowledgments} 

The authors are grateful to Yuxuan Chen for numerous helpful discussions.
MS gratefully acknowledges support for this work under ONR NICOP grant
  N62909-19-1-2128.

\bibliographystyle{unsrt}
\bibliography{references}

\end{document}